\documentclass{elsart}
\usepackage{graphicx}
\usepackage{amssymb}
\begin{document}
\begin{frontmatter}

\title
{
	Aging in citation networks
}
\author
{
Kamalika Basu Hajra and Parongama Sen}
\address{Department of Physics, University of Calcutta,
    92 Acharya Prafulla Chandra Road, Kolkata 700009, India. 
}
\begin{abstract}
In many growing networks, the age of the nodes  plays an important 
role in deciding the attachment probability of the incoming nodes. For example, in a citation network, 
very old papers are seldom cited while recent papers are usually cited
with high frequency. We study actual citation networks to find out the
distribution $T(t)$ of  $t$, the time interval
between the  published  and the cited paper. For different sets of data we find a
universal behaviour: $T(t) \sim t^{-0.9}$ for $t \leq t_c$ and $T(t) \sim t^{-2}$
for $t>t_c$ where $t_c \sim O(10)$.

PACS numbers: 05.70.Jk, 64.60.Fr, 74.40.Cx

Preprint no. CU-Physics-22/2004.
\end{abstract}
\end{frontmatter}


The question of time dependence in the attachment probability 
of the incoming nodes in a growing network has been addressed in a 
few theoretical models \cite{amaral,DM,Zhu,HS}. In these models, a new node 
gets attached to the older ones with preferential attachment
which is dependent on the degree as well as the age of the
existing node. Apart from the theoretical models, time dependence 
has also been incorporated empirically in the attachment probability 
in a model of earthquake 
network based on real data \cite{EQ}.

In the models where time dependence has been considered, the attachment
probability $\Pi(k,t)$ is generally taken to be a separable function of
the degree $k$ and age $t$ of the existing node such that

\begin{equation}
\Pi(k,t) = K(k)T(t).
\end{equation}

The functional dependence  of the attachment probability  on the degree
has been studied in quite a few real networks  
 \cite{BArev}. Based on these observations,  the $k$ dependence  
of $\Pi$  can be assumed to be proportional to 
$k^\beta$ in general \cite{KRL}, with the value of $\beta=1$ 
in most cases. However,
to the best of our knowledge, the functional form of the time dependence has not been studied 
in a similar manner for  real-world  networks.
In the theoretical models, various forms of $T(t)$  
have been considered; in \cite{amaral}, it 
has a sharp discontinuity, in \cite{Zhu} it is exponential while 
in \cite{DM} and \cite{HS},
$T(t)$ has a power law variation.

The citation network is a good example of an aging network. Here the nodes
are papers and a link is formed when one paper cites the other.   One 
can expect that in general older
papers will be cited with less probability. 
The citation network  is also  simple to model 
as older nodes cannot get new connections such that 
the evolution of the network is simply determined by the 
 links
made  by a new paper. 

We have  studied a few citation networks to find out the age dependence of
the attachment probability, or $T(t)$ of equation (1).
This study, though by no means exhaustive, is expected to give 
sufficient insight in the phenomenon of aging in networks.

The details of our study is provided below:

(a) Papers published in  a given year are chosen randomly from different databases, e.g.,  
the  databases High Energy Physics (Theory) (hep-th) and   Condensed Matter (cond-mat) Physics available at http://arxiv.org as
well as from Physical Review Letters (PRL).

(b) Suppose a paper A published in the year $t_A$  cites paper B which was published in $t_B$. 
The corresponding $t$ = $t_A-t_B$. A large number  
of $t$ values were collected with the base year, $t_A$, fixed. 
This will give us the raw distribution of the fraction of citations with 
age  $t=t_A-t_B$ which we call  $Q(t_A-t_B)$.

(c) In general, in most growing models the number of incoming nodes at a particular time
is fixed. However, the number of papers in a year is by no means fixed and this 
has to be taken care of in order to compare $T(t)$ in real and model
networks. Thus we have also studied the data $n(\tau)$ of papers published
as a function of time $\tau$ for the two  preprint archives as well as for a journal (Journal of Physics A).
In order that one can model the citation network as one in which nodes
are added one by one, 
one has to scale  $Q(t_A - t_B)$ by a scaling function $\tilde n(t_B)$ 
(related to $n(\tau)$) and identify this quantity 
as $T(t)$.



{\it {Results}} 

We have chosen 60 papers randomly from each of the databases (hep-th, cond-mat and PRL) 
belonging to a particular year (2003 for hep-th and cond-mat and 1984
for PRL). The reason for choosing these sets are that they provide 
data from different  fields of research and are also  reasonably well-separated in time (it is not
very useful 
 to go back very much in time as that would hardly 
provide data for large ages). The nature of the sets are also different in the
sense 
that the hep-th or cond-mat archives are electronic while
the other is  a printed journal. 
From the citations made in these papers the raw data $Q(t_A-t_B)$
are obtained.

\begin{center}
\begin{figure}
\includegraphics[clip,width=8cm,angle=270]{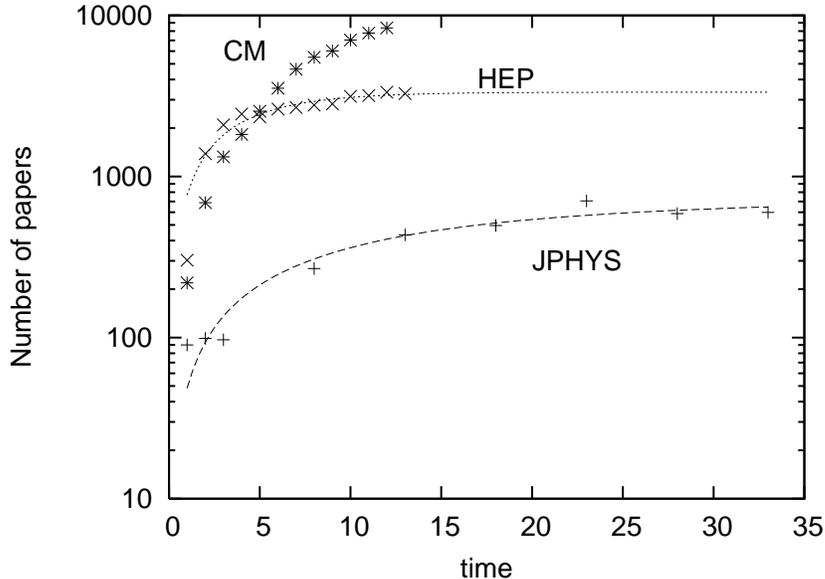}
\caption{Number of papers ($n(\tau)$) vs  time ($\tau$) plot for cond-mat (CM) 
and hep-th (HEP) arxiv and Journal of Physics A (JPHYS). While all three curves show a growth, both HEP and JPHYS curves tend to saturate. The CM curve is 
still in its growing phase. $n(\tau)\sim a(1-e^{(-b{\tau})})$ gives a reasonably good fit for HEP and JPHYS, with $a=3340$, $b=0.261$ for HEP and $a=718$, $b=0.07$ for JPHYS.}
\end{figure}
\end{center}

We next obtain the scaling function by studying the 
number of papers $n(\tau)$ published  in the three following archives:
(i) hep-th (1992-2003),
(ii) cond-mat (1993-2003)  and (iii) Journal of Physics A (JPA) (1960-2000)
in each year (as the unit of time is one year).
In Fig. 1 we have presented these  data.
The origin for each set  is taken to be the year in which the first paper  was 
published. 
As expected all the three curves show a growth, however, both the
JPA and hep-th data shows a tendency to saturate which is
not surprising. The cond-mat data appears to be still in its growing
phase. 

We assume  $n(\tau)$ to be of the 
form $a(1-\exp(-b{\tau}))$ which in fact gives reasonably good fits
with $a = 3340$, $b= 0.26$ for the hep-th data and $a= 718$, $b= 0.07$
for the JPA data.   
(We do not try to fit the cond-mat data as it is yet to reach saturation.)
The value of $b$ is quite different for the two and we choose the  value obtained
from the journal data as it is valid over a larger duration  of time
and based on papers which have actually been published.

$Q(t_A-t_B)$ is rescaled by the factor $\tilde n(t_B) = (1-e^{-0.07(t_B-t_0)})$
where $t_0$ is the "origin" in the sense that the earliest paper to be cited
was published in the year $t_0 +1$.  
Since we have kept $t_A$ fixed $\tilde n(t_B)$ can be expressed as a function of 
$t$: ~$\tilde n(t) = 1-\exp(-0.07(t_{max} -t +1)$ where $t_{max}$ is the maximum age of a cited paper in a given network. 
For example, in the PRL data, $t_A = 1984, t_{max} =116$, therefore $ t_0=1867 $
and $\tilde n(t) =1-\exp(-0.07(117-t))$.

Fig. 2 shows the scaled distribution $Q(t)/\tilde n(t)$ as a function of $t$ 
which shows very similar behaviour for all the three curves.
We notice that there are distinctly two regimes of power-law decay
of the distribution:
 $T(t)\sim t^{-\alpha_1}$ for $0 < t < t_c$
and $T(t)\sim t^{-\alpha_2}$ for $t > t_c$ where $t_c \sim O(10)$ and $\alpha_1=0.9\pm 0.1$ and $\alpha_2=2.0\pm 0.2$.\\

\begin{center}
\begin{figure}
\includegraphics[clip,width=8cm,angle=270]{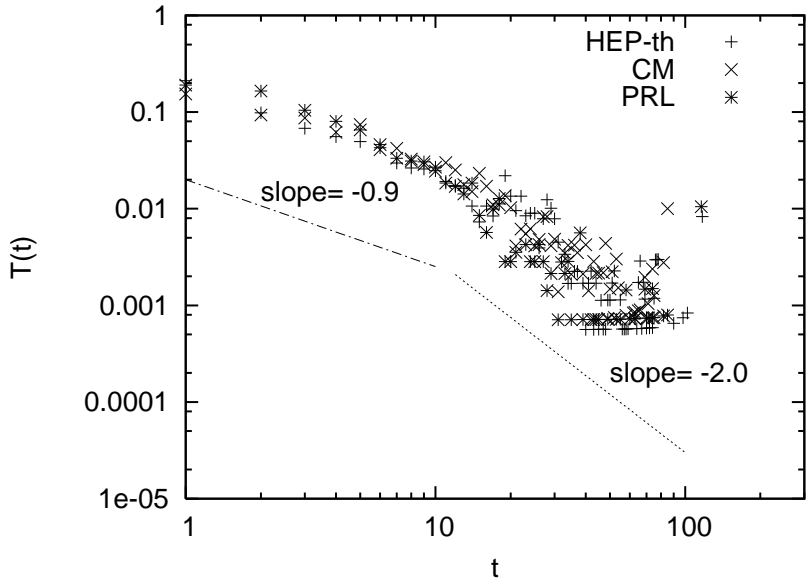}
\caption{$T(t)$  vs $t$ plot where $T(t)=Q(t)/\tilde n(t)$ is the scaled age distribution and $t$  is the age of the cited paper (refer to text for details). 
All three curves show similar behaviour, with $T(t)\sim t^{(-0.9)} $ for $0 < t < t_c$ and $f(t)\sim t^{(-2.0)}$ for $t > t_c$, where $t_c\sim O(10)$.}
\end{figure}
\end{center}

{\it Discussions:}

As mentioned earlier, the time dependence of the attachment probability can be considered 
in different forms in model networks. The present study shows that the
choice of a power law  is indeed reasonable at least for citation networks with the possibility of a crossover in the value of the exponent. 
We observe that the crossover value is roughly $t_c\sim O(10)$. From this
 it can be concluded that majority of papers have a fair chance of getting 
cited within ten years of  publication, after which fewer survive the
 'test of time'. This {\it lifespan} of ten years also signifies that most
 research problems are 
popular for such a period after which it either loses its importance or is 
replaced by newer problems or both. Hence while papers of age $\le 10$ are 
highly cited, those of age $\ge 10$ are relatively rarely cited. As an example, we indirectly found out the bulk of research devoted to persistence problems 
over time since its inception in $1993-1994$, by searching for the word 
"{\it persistence}" in the abstracts of papers submitted to the cond-mat 
archive. The percentage of such papers was $1.89$ in $1993$, increasing to a 
maximum of $2.5$ in $1997$ and then falling off gradually to $1.56$ in $2004$ 
(till July). If this trend continues, the large number of papers published 
in $1996-97$ would get much less cited after roughly ten years of their 
publication, consistent with the value of $t_c$ that we get. 

   Our sample sizes may seem rather small compared to the total size of the citation data,
but our goal has been primarily to check for the universality 
in the different subsets of citation data which we have been able to
with the chosen samples.  We have  focussed on  three kinds of subsets,
two containing papers related to a specific field  and the other to different
kinds of topics in Physics. The fact that all the three sets, very different
in nature (widely separated in subject and time), 
have  power law decay with almost the same exponents suggests that there is indeed  a  
 universal behaviour. It is expected that data from larger samples will reduce 
the fluctuations, especially 
 for $t> t_c$.

A more complete study would  of course be to find out the entire distribution
$\Pi (k,t)$ where one needs to keep track of the cumulative citations to the
cited papers and hence access to other citation databases is necessary. 
This would require longer time and more analysis and may be a topic of future research.
We believe that our study will encourage similar studies in other real networks.
It will be interesting to  find out  whether there is any universality in the form of
the age dependence factor  similar to the
degree dependence  in the preferential attachment.

Acknowledgments: KBH is grateful to CSIR (India) F.NO.9/28(609)/2003-EMR-I for financial support. PS acknowledges DST grant no.  SP/S2/M-11/99.

Email: kamalikabasu2000@yahoo.com, parongama@vsnl.net

\end{document}